\documentclass[prd,reprint,nofootinbib,showpacs,superscriptaddress]{revtex4-1}
\usepackage{graphicx} 
\usepackage{hyperref}
\usepackage{amsfonts}
\usepackage{amsmath,amssymb}
\usepackage{bm} 
\usepackage{color}
\usepackage{epstopdf} 
\usepackage{epsfig}
\usepackage{subfig} 
\usepackage{float}
\usepackage{verbatim}

{\rm }

\def\be{\begin{equation}}
 \def\ee{\end{equation}}
 \def\bea{\begin{eqnarray}}
 \def\eea{\end{eqnarray}}
 \def\bes{\begin{eqnarray}}
 \def\ees{\end{eqnarray}}
 \def\bi{\begin{itemize}}
 \def\ei{\end{itemize}} 

\def\2{\frac{1}{2}}
\def\4{\frac{1}{4}}

\begin{document}

\title{Characterizing photon number statistics using conjugate optical homodyne detection}

\author{Bing Qi}
\email{qib1@ornl.gov}
\affiliation{Quantum Information Science Group, Computational Sciences and Engineering Division,
Oak Ridge National Laboratory, Oak Ridge, TN 37831-6418, USA}
\affiliation{Department of Physics and Astronomy, The
University of Tennessee, Knoxville, TN 37996 - 1200, USA
}

\author{Pavel Lougovski}
\email{lougovskip@ornl.gov}
\affiliation{Quantum Information Science Group, Computational Sciences and Engineering Division,
Oak Ridge National Laboratory, Oak Ridge, TN 37831-6418, USA}

\author{Brian P. Williams}
\email{williamsbp@ornl.gov}
\affiliation{Quantum Information Science Group, Computational Sciences and Engineering Division,
Oak Ridge National Laboratory, Oak Ridge, TN 37831-6418, USA}

\date{\today}

\begin{abstract}

We study the problem of determining the photon number statistics of an unknown quantum state by simultaneously measuring conjugate quadratures with double homodyne detectors. Classically, the sum of the squared outputs of the two homodyne detectors is proportional to the intensity (thus the photon number) of the input light. Quantum mechanically, due to vacuum noise, the above photon number measurement is intrinsically noisy. We quantify the information gain in a single-shot measurement and discuss potential applications of this technology in quantum key distribution. We also show that the photon number statistics can be recovered in repeated measurements on an ensemble of identical input states without scanning the phase of the input state or randomizing the phase of the local oscillator used in homodyne detection.

\end{abstract}

\maketitle

\section{Introduction}
\label{sec:1}

A single photon detector (SPD) is a workhorse of modern quantum optics experiments. It is commonly used to determine the number of photons in a given light pulse in a single-shot measurement scenario \cite{Hadfield09, Eisaman11}. While photon-number-resolving detection schemes have been demonstrated using either a singls SPD or SPD arrays \cite{Rosenberg05, Waks03, Cahall17, Banaszek03, Fitch03, Zambra05}, most commercial SPDs are threshold detectors which can only distinguish between vacuum and non-vacuum states. Alternatively, given a large ensemble of identical input states, optical homodyne detection (OHD) can be used to completely reconstruct the quantum state of the ensemble, including the photon number statistics. This method is known as the optical homodyne tomography \cite{Vogel89, Smithey93, Lvovsky09}.

In OHD, a strong local oscillator (LO) is mixed with a weak input signal at a beam splitter. The interference signals can be strong enough to be detected by low cost, highly efficient photo-diodes working at room temperature. This makes OHD an appealing solution in practice, for example, in chip-size implementation \cite{Raffaelli18}. Note that the LO in OHD also functions as a mode selector: only the signals in the same spatiotemporal mode as the LO will be detected. This can be advantageous in certain applications such as quantum key distribution (QKD) \cite{Gisin02, Scarani09, Lo14, Diamanti16}, where the mode selecting function can effectively suppress broadband background noise originating from the communication channel \cite{Qi10,Heim14,Kumar15,Tobias19}.

Given the LO is sufficiently strong, a DC-balanced homodyne detector measures quadrature $X_\theta$ of the input signal, where $\theta$ is the phase of the LO \cite{Yuen83, Abbas83}. To reconstruct the full quantum state, repeated measurements are required for all values of $\theta\in [0,2\pi]$. Obviously, a crucial requirement for such a reconstruction scheme is that the phase of the input state is well controlled. However, in many applications, this requirement can not be easily satisfied. For example, in QKD, the quantum signals detected by the receiver come from a channel controlled by an adversary. In this case, we cannot make any assumptions about the phase of incoming signals. Fortunately, it has been shown that the photon number statistics of an input state with an unknown phase can still be fully recovered by using either one or two optical homodyne detectors, given the phase of the LO is uniformly randomized \cite{Munroe95, Schiller96, Leonhardt96, Richter98, Chrzanowski13}. In the case of QKD, phase randomization can be implemented by using a phase modulator driven by a random pattern, as demonstrated in \cite{Zhao07}. Nevertheless, the requirements of truly random numbers and high-speed modulator introduce additional complexities.

In this paper, we show that the conjugate optical homodyne detection scheme \cite{Richter98, Walker86, Noh91} can be used to determine the photon number statistics without controlling the phase of the input quantum state or randomizing the phase of the LO. In this scheme, two homodyne detectors are employed to measure conjugate quadratures of the input state simultaneously. We define a measurement observable $Z$ as the sum of the squared outputs of the two homodyne detectors. In classical electrodynamics, the outputs of two conjugate homodyne detectors correspond to the in-phase and out-of-phase components of an electromagnetic wave, so the observable $Z$ defined above is proportional to the intensity (or the photon number) of the input signal. In quantum mechanics, canonically conjugate quadrature components of quantum optical fields do not commute with each other and thus cannot be determined simultaneously and noiselessly due to Heisenberg's uncertainty principle. So, a single-shot measurement of $Z$ is intrinsically noisy. Nevertheless, by repeating the $Z$ measurement on a large ensemble of identical states, the photon number statistics can still be determined, as we will show in this paper. We will also study the problem of quantifying the photon number statistics of an unknown input state based on a single-shot measurement. This problem is seldom discussed in previous studies. Nevertheless, the solutions developed here may find important applications in QKD and other areas.

This paper is organized as follows: in Sec.II, we present the theory of conjugate homodyne detection. In Sec.III, we study the case of single-shot measurement and discuss its potential applications in QKD. In Sec.IV, we study the case of repeated measurements on an ensemble of identical input states. Finally, we conclude this paper with a brief summary in Sec.V.
 
\section{Conjugate homodyne detection}
\label{sec:2}

The basic setup of a conjugate homodyne detection system is shown in Fig.1. The input quantum state $\vert\psi\rangle$ is split into two parts by a symmetric beam splitter ($\textrm{BS}_1$ in Fig.1). One output of the beam splitter (mode 3 in Fig.1) is measured by an optical homodyne detector with an LO phase $\theta$; the other output state (mode 4 in Fig.1) is measured by another optical homodyne detector with an LO phase $\theta+\pi/2$. The two LOs can be generated from a common laser using a beam splitter and a $\pi/2$ phase shifter. The common phase $\theta$ is defined using the phase of the input state as a reference. Provided that the input state has an unknown phase which may change from pulse to pulse, we have no control of $\theta$ (i.e. $\theta$ is a random variable with an unknown distribution). We remark that the setup shown in Fig.1 (plus the beam splitter and the $\pi/2$ phase shifter for generating two LOs from a single laser) can be conveniently implemented with a compact commercial $90^{0}$ optical hybrid \cite{Optoplex}. In this Section, we assume noiseless homodyne detectors with unity efficiency. We will discuss the case of non-unity detection efficiency in Sec.IV.

\begin{figure}[t]
	\includegraphics[width=.45\textwidth]{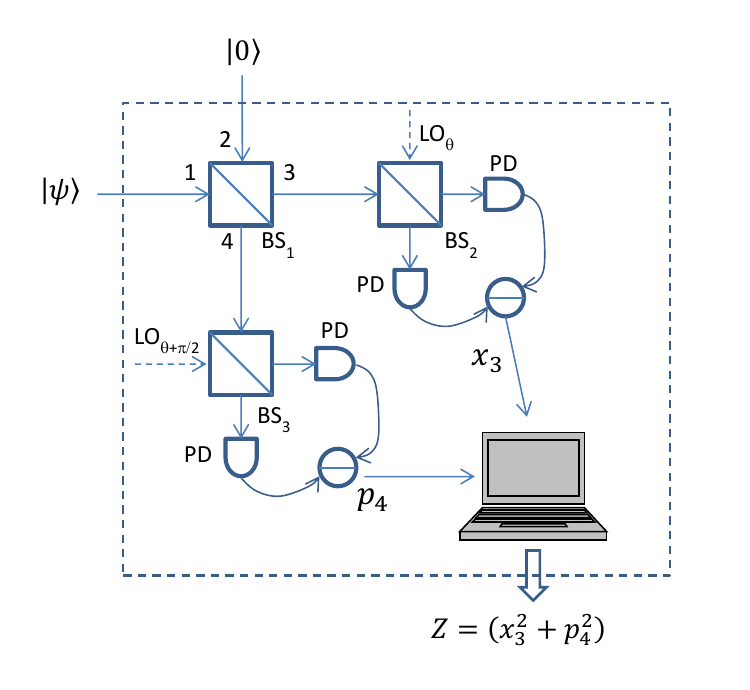}
	\captionsetup{justification=raggedright,
					singlelinecheck=false }
	\caption{Conjugate optical homodyne detection. $BS_{1-3}$: symmetric beam spliter; PD: photo detector; $LO_\theta$ ($LO_{\theta+\pi/2}$): local oscillator with phase $\theta$ ($\theta+\pi/2$).} 
	\label{fig:1}
\end{figure}

Given the LOs are sufficiently strong, the outputs of the two homodyne detectors are quadrature components of mode 3 ($X_{3,\theta}$) and mode 4 ($X_{4,\theta+\pi/2}$). For simplicity, we use $X_3$ and $X_4$ to represent $X_{3,\theta}$ and $X_{4,\theta+\pi/2}$ in the rest of the paper.

We define a new parameter  
\begin{equation}
Z=X_{3}^{2}+P_{4}^{2}
\end{equation}
as an estimation of the photon number of the input state $\vert\psi\rangle$. Intuitively, $Z$ is proportional to the intensity of the input light in classical electrodynamics.

In quantum optics, the two homodyne outputs are represented by operators $\hat{X}_{3}$ and $\hat{P}_{4}$, which are defined in terms of photon annihilation operator  $\hat{a}$ and photon creation operator $\hat{a}^{\dagger}$ as
\begin{equation}
\hat{X}_{3}=\dfrac{1}{\sqrt{2}}\left[ \hat{a}_3^{\dagger}\exp(i\theta)+\hat{a}_3\exp(-i\theta) \right], 
\end{equation}
\begin{equation}
\hat{P}_{4}=\dfrac{i}{\sqrt{2}}\left[ \hat{a}_4^{\dagger}\exp(i\theta)-\hat{a}_4\exp(-i\theta) \right].
\end{equation}

We define an operator $\hat{Z}$ as
\begin{equation}
\hat{Z}=\hat{X}_{3}^{2}+\hat{P}_{4}^{2}.
\end{equation}

Using the transformation relations of a symmetric beam splitter \cite{Loudon00}
\begin{equation}
\hat{a}_3=\dfrac{1}{\sqrt{2}}(\hat{a}_1+\hat{a}_2),
\end{equation}
\begin{equation}
\hat{a}_4=\dfrac{1}{\sqrt{2}}(\hat{a}_1-\hat{a}_2),
\end{equation}
and the commutation relation
\begin{equation}
[\hat{a}_j, \hat{a}_j^{\dagger}]=1,  j\in \left\lbrace 1,2,3,4\right\rbrace 
\end{equation}
it can be shown that
\begin{equation}\label{Eq:Eq8}
\hat{Z}=\hat{n}_1+\hat{n}_2+\hat{a}_1^{\dagger}\hat{a}_2^{\dagger}e^{i2\theta}+\hat{a}_1\hat{a}_2e^{-i2\theta}+1
\end{equation}
where $\hat{n}_1=\hat{a}_1^{\dagger}\hat{a}_1$ and $\hat{n}_2=\hat{a}_2^{\dagger}\hat{a}_2$ are photon number operators of mode 1 and 2. Obviously, $\hat{Z}$ is a Hermitian operator. Thus, it is a valid observable.

\subsection{Expectation value of $\hat{Z}$}

As shown in Fig.1, the joint input state of mode 1 and 2 is given by $\vert\psi_1 0_2\rangle$. From (8), the expectation value of $\hat{Z}$ can be determined to be
\begin{equation}\label{Eq:Eq9}
\langle\hat{Z}\rangle=\langle\psi_1 0_2\vert\hat{Z}\vert\psi_1 0_2\rangle=\langle n_1\rangle+1,
\end{equation}
where $\langle n_1\rangle$ is the average photon number of the input quantum state. The constant $1$ on the RHS of (9) can be interpreted as vacuum noise contribution \cite{Note1}. Eq.(9) shows that the average photon number of the input state can be estimated by subtracting $1$ from the expectation value of $\hat{Z}$.
 
\subsection{Variance of $\hat{Z}$}

The variance of $\hat{Z}$ is given by
\begin{equation}
\langle\Delta Z^2\rangle=\langle(\hat{Z}-\langle\hat{Z}\rangle)^2\rangle=\langle\hat{Z}^2\rangle-\langle\hat{Z}\rangle^2.
\end{equation}

From (8), it is straightforward to show
\begin{equation}\label{Eq:Eq11}
\langle\hat{Z}^2\rangle=\langle\hat{n}_1^2\rangle+3\langle\hat{n}_1\rangle+2.
\end{equation}

Using (9)-(11), we obtain
\begin{equation}
\langle\Delta Z^2\rangle=\langle\Delta n_1^2\rangle+\langle n_1\rangle+1,
\end{equation}
where $\langle\Delta n_1^2\rangle=\langle\hat{n}_1^2\rangle-\langle n_1\rangle^2$ is the variance of photon number distribution of the input state. Clearly, the additional noise of the proposed scheme is $\langle n_1\rangle+1$.

\subsection{Second-order correlation function $g^{(2)}(0)$}

The single-time second-order correlation function $g^{(2)}(0)$ is an important parameter for characterizing a photon source \cite{Glauber63}. In the case of a single homodyne detector with a phase randomized LO, it has been shown that $g^{(2)}(0)$ can be determined from the measurement statistics \cite{Roumpos13, Luders18}. Here, we show that $g^{(2)}(0)$ can also be determined from the statistics of $Z$ measurement.

As defined in \cite{Glauber63}
\begin{equation}
g^{(2)}(0)=\dfrac{\langle\hat{a}_1^{\dagger}\hat{a}_1^{\dagger}\hat{a}_1\hat{a}_1\rangle}{\langle\hat{a}_1^{\dagger}\hat{a}_1\rangle^2}.
\end{equation}

Using (7), the numerator on the RHS of (13) can be written as $
\langle\hat{a}_1^{\dagger}\hat{a}_1^{\dagger}\hat{a}_1\hat{a}_1\rangle=\langle\hat{a}_1^{\dagger}\hat{a}_1\hat{a}_1^{\dagger}\hat{a}_1-\hat{a}_1^{\dagger}\hat{a}_1\rangle=\langle\hat{n}_1^2\rangle-\langle\hat{n}_1\rangle
$. Using Eq.(\ref{Eq:Eq9}) and Eq.(\ref{Eq:Eq11}), we have
\begin{equation}
\langle\hat{a}_1^{\dagger}\hat{a}_1^{\dagger}\hat{a}_1\hat{a}_1\rangle=\langle\hat{Z}^2\rangle-4\langle\hat{Z}\rangle+2.
\end{equation}

From (9), the denominator on the RHS of (13) is simply $(\langle\hat{Z}\rangle-1)^2$. Finally, we have
\begin{equation}
g^{(2)}(0)=\dfrac{\langle\hat{Z}^2\rangle-4\langle\hat{Z}\rangle+2}{( \langle\hat{Z}\rangle-1 )^2}.
\end{equation}

\subsection{Probability density function of $\hat{Z}$}

Unlike photon number $n$, the parameter $Z$ measured in our scheme is a continuous variable. Given an arbitrary input state described by the density matrix $\rho$, we would like to determine the probability density function (PDF) of $Z$. 

The joint PDF of quadrature components $X_3$ and $P_4$ of conjugate homodyne detection has been derived in \cite{Richter98} as
\begin{multline}
P_{X_3,P_4}(x_3,p_4)=\dfrac{1}{\pi}\sum_{m,n=0}^\infty  \rho_{mn}\dfrac{\exp\left[ i(n-m)\theta\right] }{(m!n!)^{1/2}}  \\ \times (x_3-ip_4)^m(x_3+ip_4)^n \exp\left[ -(x_3^2+p_4^2)\right].   
\end{multline}

The data pair $\left(x_3,p_4\right)$ can be interpreted as the Cartesian coordinates of a point, which relates to the Polar Coordinates $\left(r,\phi \right)$ by $x_3 = r\cos\phi$ and  $p_4 = r\sin\phi$. The marginal distribution of $r$ is given by 
\begin{equation}
P_R(r) = \int_0^{2\pi} P_{X_3,P_4}(r\cos\phi,r\sin\phi) d\phi.
\end{equation}

Note the term $(x_3-ip_4)^m(x_3+ip_4)^n$ on the RHS of (16) is transformed into $r^2\exp[i(n-m)\phi]$ in the Polar Coordinates. Obviously, only terms with $n=m$ have non-zero contribution in (17). It is straightforward to show
\begin{equation}
P_R(r) = 2\exp(-r^2) \sum_{n=0}^\infty \dfrac{\rho_{nn}}{n!}r^{2n+1}.
\end{equation}

Since $Z=X_3^2+P_4^2=R^2$, the PDF $P_Z(z)$ can be determined from (18) as
\begin{equation}
P_Z(z) = \exp(-z) \sum_{n=0}^\infty \dfrac{\rho_{nn}}{n!}z^n.
\end{equation}

Eq.(19) shows the relation between the PDF of $Z$ and the photon number distribution $\rho_{nn}$ of the input state. Note that $P_Z(z)$ is only dependent on the diagonal terms of the input state, a feature we would expect from a ``phase-insensitive'' photon detector. Experimentally, Eq.(19) implies that the photon number statistics can be determined without scanning (or randomizing) the phase of the LO. 

\section{Single-shot measurement}
\label{sec:3}

In most of previous studies on OHD, one assumption is that an ensemble of identical input states are available. This allows precise quantum state characterization based on repeated measurements. However in some applications, we may need to make an estimation of photon numbers based on a single-shot measurement. One example is continuous-variable (CV) QKD \cite{Ralph99, Hillery00, GMCSQKD}, where the transmitter (Alice) encodes random bits on quadratures of either squeezed states or weak coherent states and the receiver (Bob) measures either one or two quadratures using optical homodyne detection. In these protocols, it is crucial to upper bound the photon number of the quantum state received by Bob for various reasons. First, as shown in \cite{Qin16, Qin18}, an adversary may attack Bob's QKD system by sending strong laser pulses and forcing the homodyne detectors to work in nonlinear or saturation region. It thus important to experimentally verify that the photon number of the incoming signal is within the normal range. Second, in the recent security proofs of CV-QKD with discrete modulations \cite{Ghorai19, Lin19, Kaur19}, numerical approaches are employed to determine lower bounds of secure key rates. To reduce the dimension of the relevant quantum space from infinite to finite, one important assumption is that the photon number received by Bob is finite. The scheme proposed in this paper provides a practical way to upper bound the incoming photon number. More specifically, as we will show below, given a single measurement of $Z$, it is possible to determine a threshold value of photon number $n_{th}$, such that the probability that the received signal containing $n_{th}$ or more photons is negligible.

Given the input state is a Fock state $\rho=\vert n \rangle\langle n \vert$, the likelihood of a measurement output of $z$ can be determined from (19) as
\begin{equation}
P(Z=z\vert n) = \exp(-z) \dfrac{z^n}{n!}.
\end{equation}

Using Bayes' rule the likelihood of $n$ photons coming in given the measurement output $z$ is
\begin{equation}
P(N=n\vert z) = \dfrac{P(Z=z\vert n)P_N(n)}{P_Z(z)}=\exp(-z) \dfrac{z^n}{n!}.
\end{equation}

In the lase step of (21), we have assumed that the prior $P_N(n)$ is a uniform distribution. This leads to a uniform PDF of $P_Z(z)$ from (19).

The distribution in (21) is the Poisson distribution, which quantifies the uncertainty of the photon number $n$ given a single measurement of $Z$. The uncertainty of the photon number can be determined from (21) as
\begin{equation}
\sigma=\langle\Delta n^2\rangle = z.
\end{equation}

Using (21), given a single $Z$ measurement, the probability that the received signal containing $n_{th}$ or more photons is

\begin{equation}
P(N\geq n_{th}\vert z) = \sum_{n=n_{th}}^\infty \exp(-z) \dfrac{z^n}{n!}.
\end{equation}
where $n_{th}$ is assumed to be much larger than $z$.

Using Stirling's formula $n!\geq \sqrt{2\pi}\times n^{n+\frac{1}{2}}\times\exp(-n)$, we can derive the following inequality from (23)
\begin{equation}
\begin{split}
P(N\geq n_{th}\vert z) \leq \sum_{n=n_{th}}^\infty \exp(-z) \dfrac{z^n}{\sqrt{2\pi}n^{n+\frac{1}{2}}\exp(-n)}\\
< \sum_{n=n_{th}}^\infty \exp(-z) \dfrac{z^n}{\sqrt{2\pi}n_{th}^{n}\exp(-n)} \\
= \dfrac{\exp(-z)}{\sqrt{2\pi}} \sum_{n=n_{th}}^\infty \left( \dfrac{ze}{n_{th}}\right)^n =\dfrac{\exp(-z)}{\sqrt{2\pi}} \dfrac{a^{n_{th}}}{1-a}
\end{split}
\end{equation}
where $a=\dfrac{ze}{n_{th}}$ is assumed to be much less than 1.

From (24), by choosing an appropriate $n_{th}$, $P(N\geq n_{th}\vert z)$ can be negligible. In practical QKD, the average photon number of the quantum state received by Bob is typically much less than one, so $z$ is most likely a small number. In this case, a moderate $n_{th}$ can serve as the upper bound of the incoming photon number. For example, if $z=3$, the probability that the received signal containing $n_{th}=30$ or more photons is less than $3\times 10^{-19}$.

We remark that the proposed scheme can be easily implemented in CV QKD based on conjugate homodyne detection (also called ``heterodyne detection'' in literature) \cite{Weedbrook04, Qi15}. In this case, Bob measures both X and P quadratures simultaneously for key generation. So the $Z$ information is automatically available to Bob without any changes to the QKD system or the measurement procedures.

In principle, the proposed system may also act as a threshold single-photon detector (SPD) which can discriminate vacuum state from non-vacuum states but cannot resolve photon number. This kind of detector is commonly using in discrete-variable (DV) QKD. The basic strategy is to choose an optimal threshold value $T\in\left[ 0,\infty\right)$ for the $Z$ measurement: if the measurement result $z$ is smaller (larger) than $T$, the detected quantum state is assumed to be vacuum (non-vacuum) state.

Two important parameters of a threshold SPD are detection efficiency and dark count probability. The detection efficiency $\eta$ is defined as the conditional probability that the detector reports a non-vacuum state given the input is a single photon Fock state. The dark count probability $D$ is defined as the conditional probability that the detector reports a non-vacuum state given the input is vacuum. From (20), these two parameters can be determined by
\begin{equation}
\eta = \int_{T}^{\infty} P_Z(z\vert 1) dz
\end{equation}
\begin{equation}
D = \int_{T}^{\infty} P_Z(z\vert 0) dz
\end{equation}

Fig.2 shows the simulation results of $\eta$ and $D$ as a function of the threshold value $T$. By choosing an appropriate threshold value, we can either achieve a high detection efficiency or a low dark count. Unfortunately, we cannot have both at the same time. Similar conclusions had been drawn in previous studies based on the single homodyne detection scheme \cite{Grice96}. In Fig.2, we also present the ratio $R=\eta/D$, which is an important figure of merit in applications like QKD. A state-of-the-art SPD can provide a R-value as high as $10^8$ \cite{Hadfield09}. In comparison, the R-vale of the proposed scheme is less than 10 in the region with a reasonable detection efficiency. 

\begin{figure}[t]
	\includegraphics[width=.45\textwidth]{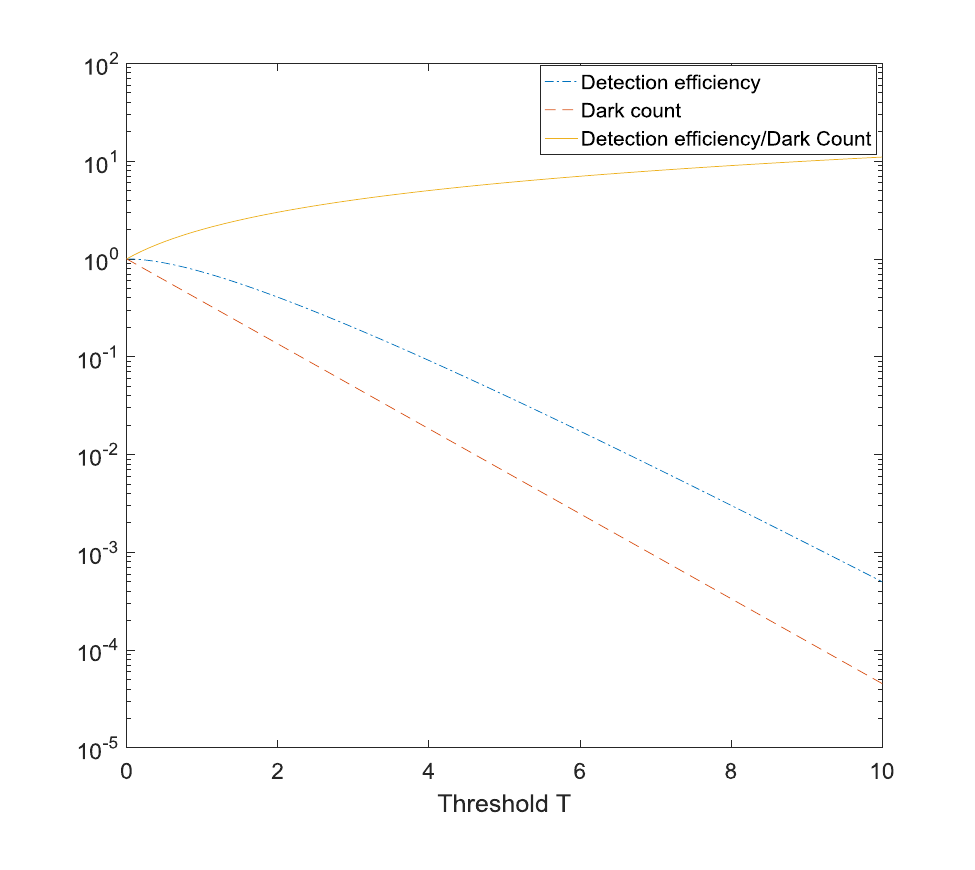}
	\captionsetup{justification=raggedright,
					singlelinecheck=false }
	\caption{(Color Online) Simulation results of detection efficiency $\eta$ (Dash-dot line), dark count probability $D$ (Dashed line), and the ratio $R=\eta/D$ (Solid line).} 
	\label{fig:2}
\end{figure}

In next section, we discuss the case of determining photon number distribution from repeated measurements.

\section{Repeated measurements}
\label{sec:4}

Given a sequence of M independent measurements of $Z$, denoted as $\{z_{i}\} = z_{1},\dots,z_{M}$, we would like to infer the underlying photon number statistics given by the distribution $p(n) = \rho_{nn}$ for $n = 0,1,\dots,N_{max}$. The observed data statistics  can be described in terms of a mixture model of $N_{max}+1$ gamma distributions ${\rm Gamma}(n+1,1)$ with mixing coefficients given by the photon number distribution $p(n)$.  In this model the conditional probability of observing an outcome $Z_{k}=z_{k}$ reads,
\begin{equation}
p(Z_{k}=z_{k}|\{p(n)\},n_{k})= p(n_{k})\frac{\exp(-z_{k})z_{k}^{n_{k}}}{n_{k}!}.
\end{equation}
The complete data likelihood function $\mathcal{L}_{c}$ for the entire measurement sequence can be written as,
\begin{equation}
\mathcal{L}_{c} = \prod_{k=1}^{M}p(Z_{k}=z_{k}|\{p(n)\},n_{k}).
\end{equation}
Here, $p(n), n = 0,\cdots,N_{max}$ are the unknown parameters (Fock state probabilities) we wish to infer and $n_{1},\dots,n_{M}$ are random variables, each in range $[0,N_{max}]$, that determine which mixture component $n_{i}\in[0,N_{max}]$ has generated an outcome $z_{i}$. Have we known the values of $n_{1},\dots,n_{M}$ then we could use the maximum-likelihood estimation (MLE) based on $\mathcal{L}_{c}$ to infer the most likely values of the parameters $p(n)$. Unfortunately the variables $n_{i},i={1,M}$ are unobserved (latent). Therefore, in order to use MLE we first need to marginalize the complete data likelihood $\mathcal{L}_{c}$ over the latent variables. The resulting marginal likelihood of the observed data reads,
\begin{equation}\label{Eq:MarginalLikelihood}
\mathcal{L} = \prod_{k=1}^{M}\sum_{n_{k}=0}^{N_{max}}p(Z_{k}=z_{k}|\{p(n)\},n_{k}),
\end{equation}
where we used the following convention,
\begin{equation}
\prod_{k=1}^{M}\sum_{n_{k}=0}^{N_{max}}=\sum_{n_{1}=0}^{N_{max}}\cdots\sum_{n_{M}=0}^{N_{max}}.
\end{equation}

In principle one now can try and maximize the function $\mathcal{L}$ with respect to parameters $p(n)$, but in practice the complexity of this task will grow exponentially with the number of measurements $M$. Fortunately, there is a way to determine the maximum of $\log\mathcal{L}$ that avoids explicit maximization of $\mathcal{L}$. This method is widely used for inference in mixture models and is called expectation-maximization (EM) algorithm~\cite{EM}. Applicability of EM in the context of homodyne measurement of photon statistics has been advocated previously~\cite{Banaszek98}. EM is an iterative procedure that uses the expected value of the complete data likelihood $\mathcal{L}_{c}$ with respect to the latent random variables $n_{1},\dots,n_{M}$ as a maximization objective function for determining $p^{t}(n)$ -- the $t$-th iteration estimate of the parameters $p(n)$. Here is how it works:\\
~\\
\noindent First, set initial estimates of the parameters $p^{0}(n)$ to some values. We chose a uniform prior
\begin{equation}p^{0}(n)=\frac{1}{N_{max}+1}\nonumber\end{equation}
as it is uninformative and easy to implement. Due to the multimodality of Eq. (\ref{Eq:MarginalLikelihood}) the choice of prior will bias the EM reconstruction given below. It is an open problem to identify the best prior for this application\footnote{In principal, one can choose any set of positive real numbers such that $\sum\limits_{n=0}^{N_{max}}p^{0}(n)=1$.}. Next, repeat the following steps until convergence criteria are satisfied.
\begin{itemize}
    \item At the $t$-th iteration, update probabilities $p(n_{k}|z_{k},\{p^{t}(n)\})$ for all latent variables $n_{1},\dots,n_{M}$ using Bayes rule with $p^{t}(n)$ as a prior,\vspace*{0.5cm}\\
    \hspace*{0.4cm}$p(n_{k}|z_{k},\{p^{t}(n)\}) = \frac{p(Z_{k}=z_{k}|\{p^{t}(n)\},n_{k})}{\sum\limits_{n_{k}=0}^{N_{max}}p(Z_{k}=z_{k}|\{p^{t}(n)\},n_{k})}$
    \item Calculate the expected value of the complete data $\log$ likelihood with respect to the updated distribution $p(n_{k}|z_{k},\{p^{t}(n)\})$,\vspace*{0.5cm}\\
    \hspace*{0.4cm}$Q(\{p(n)\}|\{p^{t}(n)\}) = {\rm E}_{\{n_{k}\}|\{z_{k}\},\{p^{t}(n)\}}[\log \mathcal{L}_{c}]=$\\
    $\sum\limits_{k=1}^{M}\sum\limits_{j=0}^{N_{max}}p(j|z_{k},\{p^{t}(n)\})\log[p(Z_{k}=z_{k}|\{p(n)\},j)]$
    \item Find parameter values $\{p_{max}(n)\}$ that maximize $Q(\{p(n)\}|\{p^{t}(n)\})$ and set the next iteration estimates of photon number probabilities $\{p^{t+1}(n)\}=\{p_{max}(n)\}$. Note that the values of $\{p_{max}(n)\}$ can be calculated analytically,
    \vspace*{0.5cm}\\
    \hspace*{1.1cm}$p^{t+1}(n) = \frac{1}{M}\sum\limits_{j=1}^{M}p(n|z_{j},\{p^{t}(n)\})$.
\end{itemize}
It can be shown~\cite{EM} that iterative maximization of $Q(\{p(n)\}|\{p^{t}(n)\})$ also results in the maximization of the marginal likelihood $\mathcal{L}$ in Eq.(\ref{Eq:MarginalLikelihood}). Therefore, after a sufficient number of iterations $t_s$ our MLE estimator of the photon number statistics is given by the distribution $\{p^{t_{s}}(n)\}$. 

\subsection{Simulation results}

To illustrate our EM-based photon number statistics inference scheme we applied it to a sequence of $32768$ simulated homodyne measurement outcomes for a coherent state $\rho = |\alpha\rangle\langle\alpha|$ with the mean photon number $|\alpha|^{2}=5$. To reconstruct the photon number statistics from the simulated measurement data, we have selected a mixture model with $N_{max}=20$ ($21$-component model). The results of EM reconstruction (after 9 iterations) are plotted on Fig.3 and demonstrate a good quantitative agreement with the true state.

\begin{figure}[t]
\includegraphics[width=.45\textwidth]{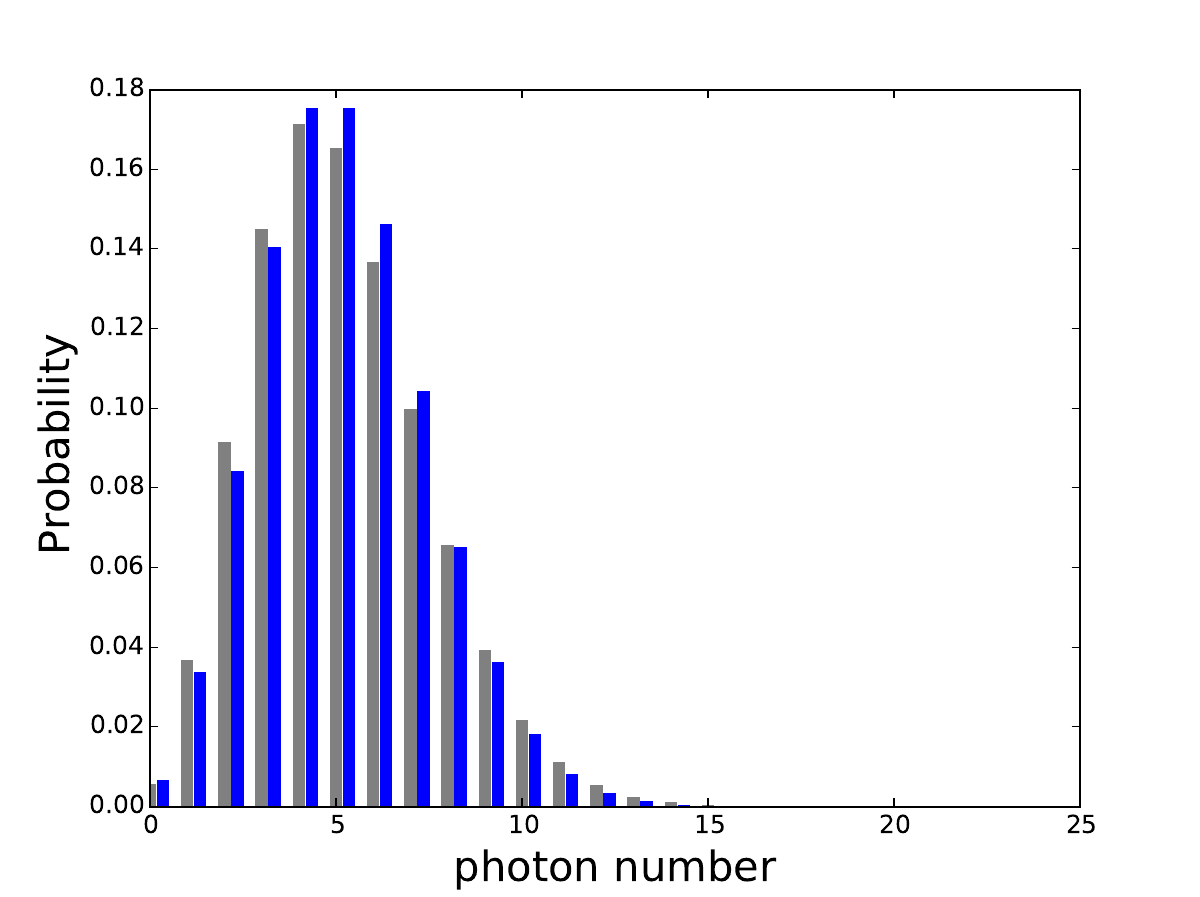}
\captionsetup{justification=raggedright}
\caption{(Color Online) EM reconstruction of photon number statistics from a sequence of simulated homodyne measurements for a coherent state. The blue histogram bars represent a true photon distribution. The gray histogram bars correspond to a distribution reconstructed by using the EM algorithm.}\label{Fig:3}
\end{figure}

\subsection{Experimental results}

We apply the theory described above in experiments to reconstruct photon number distributions of a weak coherent state and a thermal state.

In Sec.II, we have assumed perfect homodyne detectors with unity efficiency. Here we consider practical detector with non-unity detection efficiency. It is well known that a realistic photo-detector with efficiency $\eta$ can be modeled by placing a virtual beam splitter (with a transmittance of $\eta$) in front of an ideal detector \cite{Yuen78}. By assuming the four photo-detectors have identical efficiency, we can model the conjugate homodyne detection using the setup shown in Fig.4(a). In Appendix A we will show that given the LOs are strong enough, the setup in Fig.4(a) is equivalent to that in Fig.4(b), where the four virtual beam splitters in front of the photo-detectors are replaced by a common virtual beam splitter (with the same transmittance) at the input of the first beam splitter. This is convenient in practice since we can apply the theory in Sect. II directly to the experimental results by assuming the photo-detectors are ideal. The photon number distribution reconstructed this way is related to that of the input state by the Bernoulli transformation \cite{Campos89}. By further applying the inverse Bernoulli transformation, the photon number distribution of the input state can be determined.

In our experiments, either a weak coherent source or a thermal source is employed as the input. The state after the virtual beam splitter in Fig.4(b) is still a coherent state (or a thermal state) with a reduced average photon number. This is convenient since we can simply redefine the state after the virtual beam splitter as the input state and assume the detection efficiency is one.   

\begin{figure}[t]
	\includegraphics[width=.45\textwidth]{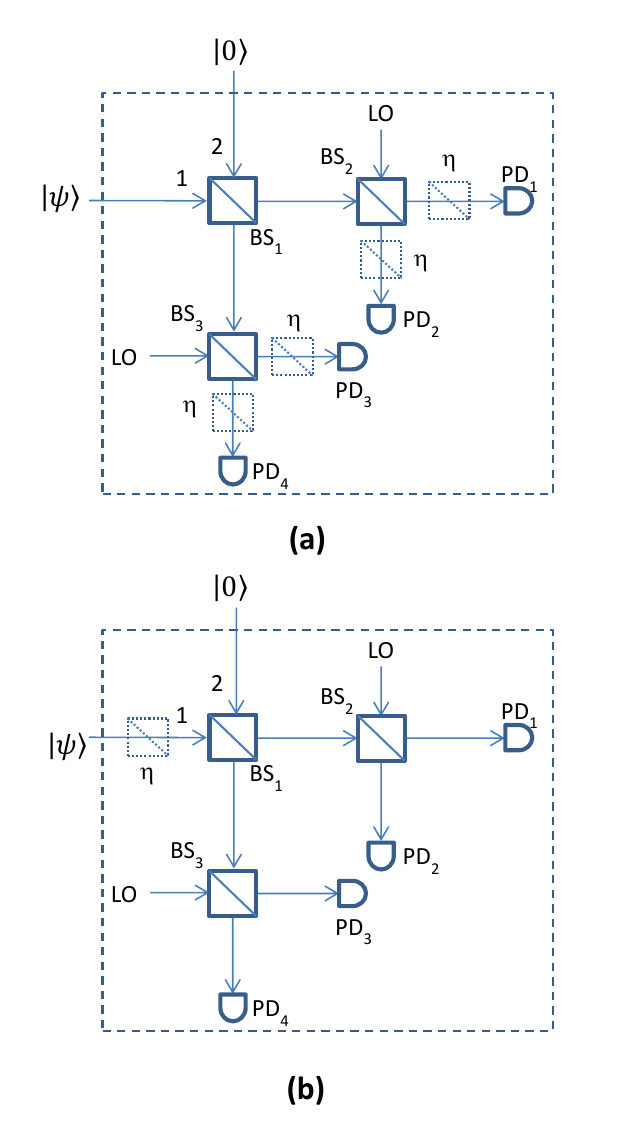}
	\captionsetup{justification=raggedright,
					singlelinecheck=false }
	\caption{Models of realistic photo-detector with detection efficiency $\eta$. (a) The actual setup. (b) An equivalent model of (a). See details in Appendix A.} 
	\label{fig:4}
\end{figure}

The experimental setup is shown in Fig.5. A 1550nm continuous wave (CW) laser is employed as the LO. The conjugate optical homodyne detection system is constructed by a commercial $90^o$ optical hybrid (Optoplex) and two $350$ MHz balanced amplified photodetectors (Thorlabs). Variable optical attenuators are used to balance the detection efficiency of different channels and control the average photon number of the input state. The outputs of the two balanced photodetectors are sampled by a two-channel data acquisition board (Texas Instruments). The overall efficiency of the detection system is 0.5, with electrical noises variances (in shot-noise unit) of $\sigma^{2}_{X_3} = 0.21, \sigma^{2}_{P_4} = 0.16$ for the quadratures $X_{3}$ and $P_{4}$ respectively.

\begin{figure}[t]
	\includegraphics[width=.45\textwidth]{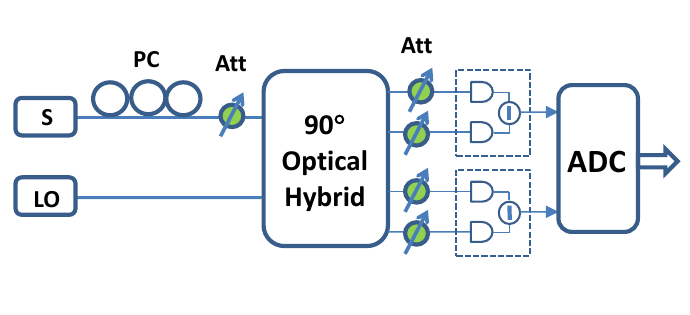}
	\captionsetup{justification=raggedright,
					singlelinecheck=false }
	\caption{Experimental setup. S-photon source under test; LO-local oscillator; PC-polarization controller; Att-variable optical attenuator; ADC-data acquisition board. } 
	\label{fig:5}
\end{figure}

In the first experiment, a heavily attenuated laser source is used to provide a weak coherent state input. By adjusting the variable optical attenuator, the average photon number within one sampling window has been set to 5 (after correction of detection efficiency). In the second experiment, an amplified spontaneous emission (ASE) source is used to provide a thermal state input (with an average photon number of 15.3). Note, while the output of the ASE source contains multiple modes, the homodyne detector selectively measures the one matched with the LO mode. Limited by the memory size of the data acquisition board, 32728 quadrature pairs are sampled in each measurement.

Using (15), the $g^{(2)}(0)$ factors of the two sources have been determined to be $1.11\pm0.02$ (weak coherent source) and $1.94\pm0.02$ (thermal source) correspondingly, where the uncertainty quantifies statistic fluctuation. These results match with the theoretical values of 1 (ideal coherent state) and 2 (ideal thermal state) reasonably well.

We apply the EM algorithm described above to the experimental data in order to reconstruct the photon number statistics of the two light sources. In Fig.6 we plot reconstruction results for a weak coherent state. The blue bars represent the photon number statistic of a coherent state $|\alpha\rangle$ with the mean number of photons $|\alpha|^2 = 5$, assuming noiseless detectors. The red bars correspond to the numerically synthesized data from the coherent state $|\alpha\rangle$ with added Gaussian noise which mimics the effect of noisy photo detectors. The green bars depict the photon number statistics obtained from raw experimental data by using the EM algorithm. We note that the red and green histograms look remarkably similar which implies that experimental data come from a coherent state affected by the detector noise. In Fig.7 we depict EM reconstruction results for a thermal state with a mean photon number of 15.3. The black line represents the theoretical distribution of an ideal thermal state. The red bar are EM reconstruction results using numerically simulated quadrature measurement data with added detector noise. Finally, the blue bars correspond to EM reconstruction of actual experimental data. We notice that for the thermal state the three distributions are very close visually. This is because the detector noise is much smaller than the mean number of photons in this case. Therefore, the noise effects are not as pronounced as in the case on a weak coherent state.      
\begin{figure}[t]
	\includegraphics[width=.45\textwidth]{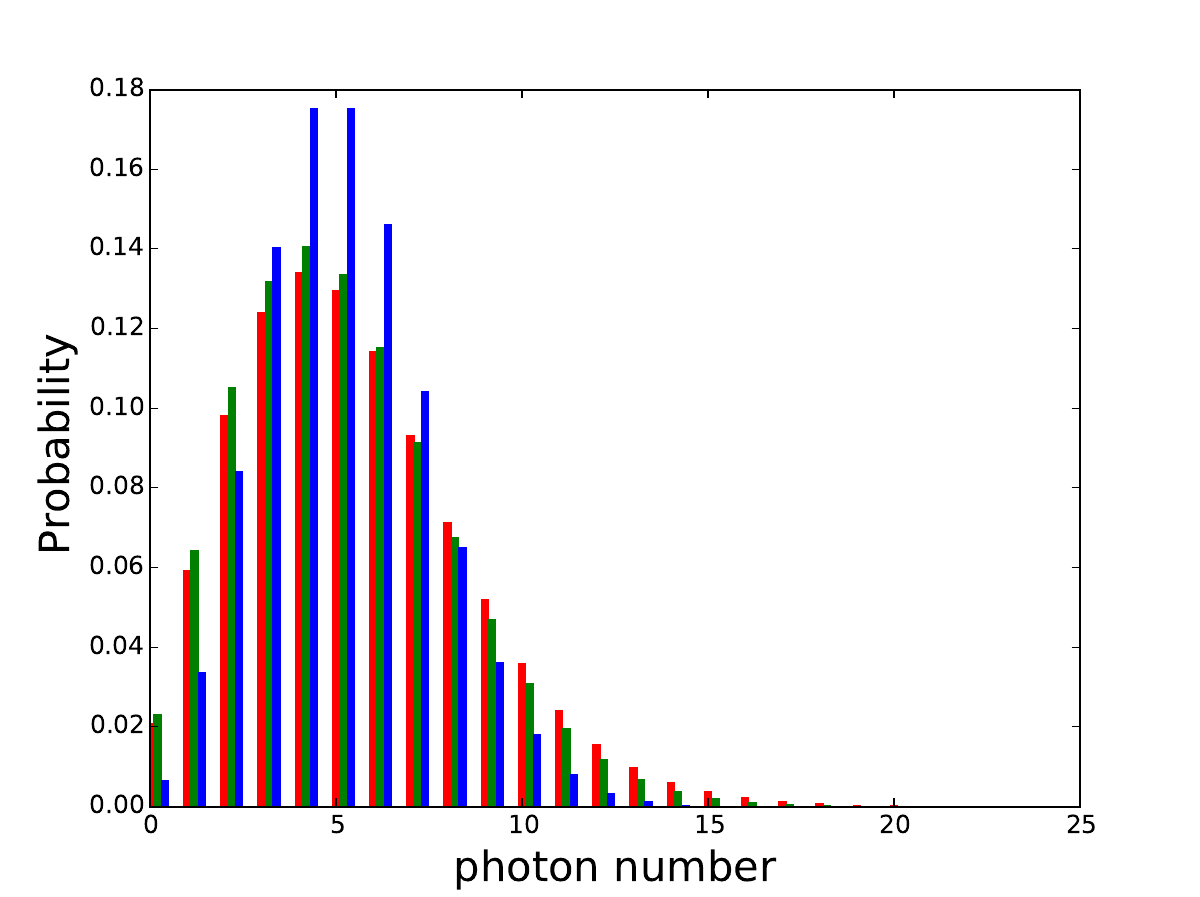}
	\captionsetup{justification=raggedright,
					singlelinecheck=false }
	\caption{(Color Online) Histograms of the reconstructed photon number distributions for a coherent state $|\alpha\rangle$: a) Photon number statistics for $|\alpha|^2=5$ (blue) b) Simulated measurement for $|\alpha|^2=5$ data with added detector noise (red) c) Experimental data (green).} 
	\label{fig:6}
\end{figure}

\begin{figure}[t]
	\includegraphics[width=.45\textwidth]{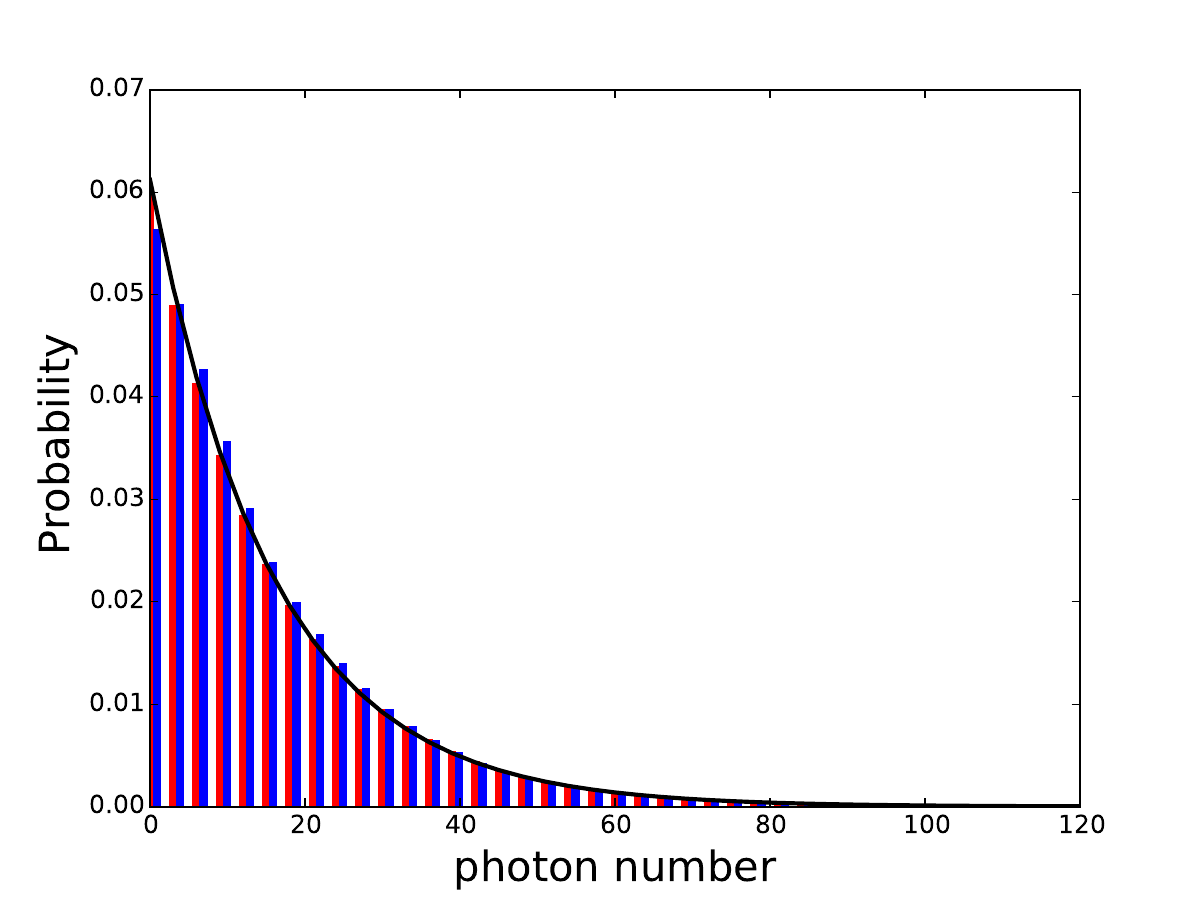}
	\captionsetup{justification=raggedright,
					singlelinecheck=false }
	\caption{(Color Online) Histograms of the reconstructed photon number distributions for a thermal state with the mean photon number $=15.3$: a) Simulated measurement data with added detector noise (red) b) Reconstruction results for experimental data  (blue) c) Thermal distribution with the mean photon number $=15.3$ (black).} 
	\label{Fig:7}
\end{figure}

\section{Summary}
\label{sec:5}

Classically, the intensity of a single-mode light pulse can be determined by measuring two conjugate quadratures simultaneously. Quantum mechanically, the above measurement process is intrinsically noisy. In this paper, we develop theoretical tools to reconstruct photon number statistics of a single-mode quantum state by performing conjugate homodyne detection. Comparing with previous studies based on single homodyne detection, no LO phase randomization is required in our scheme. We also show how to determine an upper bound of incoming photon number based on a single-shot measurement. Such a technology could be useful in CV-QKD to prevent bright pulse attack or validate crucial assumptions in security proofs.

We acknowledge helpful comments from Ryan S. Bennink, Warren Grice, Charles C. W. Lim and Nicholas A. Peters. This manuscript has been authored by UT-Battelle, LLC under Contract No. DE-AC05-00OR22725 with the U.S. Department of Energy. The United States Government retains and the publisher, by accepting the article for publication, acknowledges that the United States Government retains a non-exclusive, paid-up, irrevocable, world-wide license to publish or reproduce the published form of this manuscript, or allow others to do so, for United States Government purposes. The Department of Energy will provide public access to these results of federally sponsored research in accordance with the DOE Public Access Plan (http://energy.gov/downloads/doe-public-access-plan). The authors acknowledge support from ORNL laboratory directed research and development program.

\appendix

\section{Detector efficiency in the conjugate homodyne scheme}

In this Appendix, we will show that given the LOs are strong enough, the setup shown in Fig.4(a) is equivalent to that in Fig.4(b), where the four virtual beam splitters in front of the photo-detectors are replaced by a common virtual beam splitter (with the same transmittance) at the input path of the first beam splitter.

Given the LO is strong enough, a single DC-balanced homodyne detector using two realistic (non-unity efficiency) and identical photo-detectors can be modeled by the one with ideal photo-detectors by placing a virtual beam splitter at the signal input \cite{Leonhardt93}. This allows us to replace the four virtual beam splitters in Fig.4(a) by two virtual beam splitters with the same transmittance (one at each of the output port of the first beam splitter). We will show that these two virtual beam splitters can be further replaced by one as shown in Fig.4(b). More specifically, we will show the two models in Fig.8 are equivalent to each other.

\begin{figure}[t]
	\includegraphics[width=.45\textwidth]{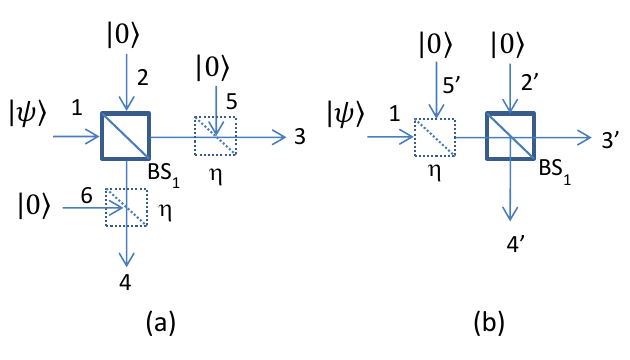}
	\captionsetup{justification=raggedright,
					singlelinecheck=false }
	\caption{Two equivalent models. (a) A symmetric beam splitter followed by two virtual beam splitters. (b) A virtual beam splitter placed in front of the symmetric beam splitter.} 
	\label{fig:8}
\end{figure}

For simplicity, we define the transmittance of the virtual beam splitter as $\eta=\cos\gamma$.

From Fig.8(a) and using the transmission relations of lossless beam splitter, we have
\begin{multline}
\hat{X}_{3}=\dfrac{1}{\sqrt{2}}\cos\gamma\hat{X}_{1}+\dfrac{1}{\sqrt{2}}\cos\gamma\hat{X}_{2}+\sin\gamma\hat{X}_{5} \\ =\dfrac{1}{\sqrt{2}}\cos\gamma\hat{X}_{1}+\dfrac{1}{\sqrt{2}}\sqrt{1+\sin^2\gamma}\hat{X}_{V1}
\end{multline}
where $\hat{X}_{V1}=\dfrac{\cos\gamma}{\sqrt{1+\sin^2\gamma}}\hat{X}_{2}+\dfrac{\sqrt{2}\sin\gamma}{\sqrt{1+\sin^2\gamma}}\hat{X}_{5}$. Since the inputs of mode 2 and mode 5 in Fig.8(a) are vacuum, the unitary transformation described above yields another vacuum state. So $\hat{X}_{V1}$ at the RHS of (A1) can be interpreted as the X-quadrature of vacuum state.

Similarly, the P-quadrature of mode 4 in Fig.8(a) is given by 
\begin{multline}
\hat{P}_{4}=\dfrac{1}{\sqrt{2}}\cos\gamma\hat{P}_{1}-\dfrac{1}{\sqrt{2}}\cos\gamma\hat{P}_{2}+\sin\gamma\hat{P}_{5} \\ =\dfrac{1}{\sqrt{2}}\cos\gamma\hat{P}_{1}+\dfrac{1}{\sqrt{2}}\sqrt{1+\sin^2\gamma}\hat{P}_{V2}
\end{multline}
where $\hat{P}_{V2}=-\dfrac{\cos\gamma}{\sqrt{1+\sin^2\gamma}}\hat{P}_{2}+\dfrac{\sqrt{2}\sin\gamma}{\sqrt{1+\sin^2\gamma}}\hat{P}_{6}$.

We can apply the same process in the model shown in Fig.8(b) and have the following relations:
\begin{multline}
\hat{X}_{3'}=\dfrac{1}{\sqrt{2}}\cos\gamma\hat{X}_{1}+\dfrac{1}{\sqrt{2}}\sin\gamma\hat{X}_{5'}+\dfrac{1}{\sqrt{2}}\hat{X}_{2'} \\ =\dfrac{1}{\sqrt{2}}\cos\gamma\hat{X}_{1}+\dfrac{1}{\sqrt{2}}\sqrt{1+\sin^2\gamma}\hat{X}_{V3}
\end{multline}
where $\hat{X}_{V3}=\dfrac{\sin\gamma}{\sqrt{1+\sin^2\gamma}}\hat{X}_{5'}+\dfrac{1}{\sqrt{1+\sin^2\gamma}}\hat{X}_{2'}$
\begin{multline}
\hat{P}_{4'}=\dfrac{1}{\sqrt{2}}\cos\gamma\hat{P}_{1}+\dfrac{1}{\sqrt{2}}\sin\gamma\hat{P}_{5'}-\dfrac{1}{\sqrt{2}}\hat{P}_{2'} \\ =\dfrac{1}{\sqrt{2}}\cos\gamma\hat{P}_{1}+\dfrac{1}{\sqrt{2}}\sqrt{1+\sin^2\gamma}\hat{P}_{V4}
\end{multline}
where $\hat{P}_{V4}=\dfrac{\sin\gamma}{\sqrt{1+\sin^2\gamma}}\hat{P}_{5'}-\dfrac{1}{\sqrt{1+\sin^2\gamma}}\hat{P}_{2'}$.

It is easy to show $X_{V1}$, $P_{V2}$, $X_{V3}$, $P_{V4}$ are independent and identically distributed random variables. From (A1)-(A4), the joint probability of $X_3$ and $P_4$ is the same as that of $X_{3'}$ and $P_{4'}$. So the two models given in Fig.8 are equivalent.


\begin{thebibliography}{00}

\bibitem{Hadfield09}
R. H. Hadfield, Single-photon detectors for optical quantum information applications, Nature Photonics \textbf{3}, 696 (2009).

\bibitem{Eisaman11}
M. D. Eisaman, J. Fan, A. Migdall, and S. V. Polyakov, Single-Photon Sources and Detectors, Rev. Sci. Instrum. \textbf{82}, 071101 (2011).

\bibitem{Rosenberg05}
D. Rosenberg, A. E. Lita, A. J. Miller, and S. W. Nam, Noise-free high-efficiency photon-number-resolving detectors, Phys. Rev. A \textbf{71}, 061803(R) (2005).

\bibitem{Waks03}
E. Waks, K. Inoue, W. D. Oliver, E. Diamanti, and Y. Yamamoto, High-efficiency photon-number detection for quantum information processing, J. Sel. Top. Quantum Electron. \textbf{9}, 1502 (2003). 

\bibitem{Cahall17}
C. Cahall, K. L. Nicolich, N. T. Islam, G. P. Lafyatis, A. J. Miller, D. J. Gauthier, and J. Kim, Photon detection using a conventional superconducting nanowire single-photon detector, Optica \textbf{4}, 1534 (2017).

\bibitem{Banaszek03}
K. Banaszek and I. A. Walmsley, Photon counting with a loop detector, Opt. Lett. \textbf{28}, 52 (2003). 

\bibitem{Fitch03}
M. J. Fitch, B. C. Jacobs, T. B. Pittman, and J. D. Franson, Photon-number resolution using time-multiplexed single-photon detectors, Phys. Rev. A \textbf{68}, 043814 (2003). 

\bibitem{Zambra05}
G. Zambra, A. Andreoni, M. Bondani, M. Gramegna, M. Genovese, G. Brida, A. Rossi, and M. G. A. Paris, Experimental reconstruction of photon statistics without photon counting, Phys. Rev. Lett. \textbf{95}, 063602 (2005).


\bibitem{Vogel89}
K. Vogel and H. Risken, Determination of quasiprobability distributions in terms of probability distributions for the rotated quadrature phase, Phys. Rev. A \textbf{40}, 2847 (1989).

\bibitem{Smithey93}
D. T. Smithey, M. Beck, M. G. Raymer, and A. Faridani, Measurement of the Wigner Distribution and the Density Matrix of a Light Mode Using Optical Homodyne Tomography: Application to Squeezed States and the Vacuum, Phys. Rev. Lett. \textbf{70}, 1244 (1993).

\bibitem{Lvovsky09}
A. I. Lvovsky and M. G. Raymer, Continuous-variable optical quantum-state tomography, Rev. Mod. Phys. \textbf{81}, 299 (2009).

\bibitem{Raffaelli18}
F. Raffaelli, G. Ferranti, D. H. Mahler, P. Sibson, J. E. Kennard, A. Santamato, G. Sinclair, D. Bonneau, M. G. Thompson, and J. C. F. Matthews, A homodyne detector integrated onto a photonic chip for measuring quantum states and generating random numbers, Quantum Sci. Technol. \textbf{3}, 025003 (2018).

\bibitem{Gisin02}
N. Gisin, G. Ribordy, W. Tittel, and H. Zbinden, Quantum cryptography, Rev. Mod. Phys. \textbf{74}, 145 (2002).

\bibitem{Scarani09}
V. Scarani, H. Bechmann-Pasquinucci, N. J. Cerf, M. Du\v{s}ek, N. L\"{u}tkenhaus, and M. Peev, The security of practical quantum key distribution, Rev. Mod. Phys. \textbf{81}, 1301 (2009).

\bibitem{Lo14}
H.-K. Lo, M. Curty, and K. Tamaki, Secure quantum key distribution, Nature Photonics \textbf{8}, 595 (2014).

\bibitem{Diamanti16}
E. Diamanti, H.-K. Lo, B. Qi, and Z. Yuan, Practical challenges in quantum key distribution, npj Quantum Information \textbf{2}, 16025 (2016).

\bibitem{Qi10}
B. Qi, W. Zhu, L. Qian, and H.-K. Lo, Feasibility of quantum key distribution through a dense wavelength division multiplexing network, New J. Phys. \textbf{12}, 103042 (2010).

\bibitem{Heim14}
B. Heim, C. Peuntinger, N. Killoran, I. Khan, C. Wittmann, Ch. Marquardt, and G. Leuchs, Atmospheric continuous-variable quantum communication, New J. Phys. \textbf{16}, 113018 (2014).

\bibitem{Kumar15}
R. Kumar, H. Qin, and R. All\'{e}aume, Coexistence of continuous variable QKD with intense DWDM classical channels, New J. Phys. \textbf{17}, 043027 (2015).

\bibitem{Tobias19}
A. Tobias, et al., Wavelength division multiplexing of continuous variable quantum key distribution and 18.3 Tbit/s data channels, Communications Physics \textbf{2}, 9 (2019).

\bibitem{Yuen83}
H. P. Yuen and V. W. S. Chan, Noise in homodyne and heterodyne detection, Opt. Lett. \textbf{8}, 177 (1983).

\bibitem{Abbas83}
G. L. Abbas, V. W. S. Chan, and T. K. Yee, Local-oscillator excess-noise suppression for homodyne and heterodyne detection, Opt. Lett. \textbf{8}, 419 (1983).

\bibitem{Munroe95}
M. Munroe, D. Boggavarapu, M. E. Anderson, and M. G. Raymer, Photon-number statistics from the phase-averaged quadrature-field distribution: Theory and ultrafast measurement, Phys. Rev. A \textbf{52}, R924 (1995).

\bibitem{Schiller96}
S. Schiller, G. Breitenbach, S. F. Pereira, T. M\"{u}ller, and J. Mlynek, Quantum Statistics of the Squeezed Vacuum by Measurement of the Density Matrix in the Number State Representation, Phys. Rev. Lett. \textbf{77}, 2933 (1996).

\bibitem{Leonhardt96}
U. Leonhardt, M. Munroe, T. Kiss, Th. Richter, and M. G. Raymer, Sampling of photon statistics and density matrix using homodyne detection, Opt. Commun. \textbf{127}, 144 (1996).

\bibitem{Richter98}
Th. Richter, Determination of photon statistics and density matrix from double homodyne detection measurements, J. Mod. Opt. \textbf{45}, 1735 (1998).

\bibitem{Chrzanowski13}
H. M. Chrzanowski, S. M. Assad, J. Bernu, B. Hage, A. P. Lund, T. C. Ralph, P. K. Lam and T. Symul, Reconstruction of photon number conditioned states using phase randomized homodyne measurements, J. Phys. B: At. Mol. Opt. Phys. \textbf{46}, 104009 (2013).

\bibitem{Zhao07}
Y. Zhao, B. Qi, and H.-K. Lo, Experimental quantum key distribution with active phase randomization, Appl. Phys. Lett. \textbf{90}, 044106 (2007).

\bibitem{Walker86}
N. G. Walker and J. E. Carroll, Multiport homodyne detection near the quantum noise limit, Opt. Quantum Electron. \textbf{18}, 355 (1986).

\bibitem{Noh91}
J. W. Noh, A. Foug\`{e}res, and L. Mandel, Measurement of the Quantum Phase by Photon Counting, Phys. Rev. Lett. \textbf{67}, 1426 (1991).

\bibitem{Optoplex}
See, for example, http://www.optoplex.com/

\bibitem{Loudon00}
R. Loudon, \emph{The Quantum Theory of Light} (Oxford University Press, Oxford, England, 2000).

\bibitem{Note1}
We could remove the constant $1$ in (9) by redefining (1) as $Z=X_{3}^{2}+P_{4}^{2}-1$. However, such a definition may result a negative measurement result of $Z$ in the case of single-shot measurement.

\bibitem{Glauber63}
R. J. Glauber, The Quantum Theory of Optical Coherence, Phys. Rev. \textbf{130}, 2529 (1963).

\bibitem{Roumpos13}
G. Roumpos and S. T. Cundiff, Photon number distributions from a diode laser, Opt. Lett. \textbf{38}, 139 (2013).

\bibitem{Luders18}
C. L\"{u}ders, J. Thewes, and M. Assmann, Real time $g^{(2)}$ monitoring with 100 kHz sampling rate, Optics express \textbf{26}, 24854-24863 (2018).

\bibitem{Ralph99}
T. C. Ralph, Continuous variable quantum cryptography, Phys. Rev. A \textbf{61}, 010303(R) (1999).

\bibitem{Hillery00}
M. Hillery, Quantum cryptography with squeezed states, Phys. Rev. A \textbf{61}, 022309 (2000).

\bibitem{GMCSQKD}
F. Grosshans, G. V. Assche, J. Wenger, R. Brouri, N. J. Cerf, and Ph. Grangier, Quantum key distribution using gaussian-modulated coherent states, Nature \textbf{421}, 238 (2003).

\bibitem{Qin16}
H. Qin, R. Kumar, and R. All\'{e}aume, Quantum hacking: Saturation attack on practical continuous-variable quantum key distribution, Phys. Rev. A \textbf{94}, 012325 (2016).

\bibitem{Qin18}
H. Qin, R. Kumar, V. Makarov, and R. All\'{e}aume, Homodyne-detector-blinding attack in continuous-variable quantum key distribution, Phys. Rev. A \textbf{98}, 012312 (2018).

\bibitem{Ghorai19}
S. Ghorai, P. Grangier, E. Diamanti, and A. Leverrier, Asymptotic security of continuous-variable quantum key distribution with a discrete modulation, Physical Review X \textbf{9}, 021059 (2019).

\bibitem{Lin19}
J. Lin, T. Upadhyaya, and N. L\"{u}tkenhaus, Asymptotic security analysis of discrete-modulated continuous-variable quantum key distribution, arXiv:1905.10896 (2019).

\bibitem{Kaur19}
E. Kaur, S. Guha, and M. M. Wilde, Asymptotic security of discrete-modulation protocols for continuous-variable quantum key distribution, arXiv:1901.10099 (2019).

\bibitem{Weedbrook04}
C. Weedbrook, A.M. Lance, W.P. Bowen, T. Symul, T. C. Ralph, and P. K. Lam, Quantum Cryptography Without Switching, Phys. Rev. Lett. \textbf{93}, 170504 (2004).

\bibitem{Qi15}
B. Qi, P. Lougovski, R. Pooser, W. Grice, and M. Bobrek, Generating the Local Oscillator ``Locally" in Continuous-Variable Quantum Key Distribution Based on Coherent Detection, Phys. Rev. X \textbf{5}, 041009 (2015).

\bibitem{Grice96}
W. Grice and I. A. Walmsley, Homodyne Detection in a Photon Counting Application, J. Mod. Opt. \textbf{43}, 795 (1996).

\bibitem{EM}
A. P. Dempster, N. M. Laird and D. B. Rubin, Maximum Likelihood from Incomplete Data via the EM Algorithm, Journal of the Royal Statistical Society. Series B (Methodological) \textbf{39}, pp. 1-38 (1977).

\bibitem{Banaszek98}
K. Banaszek, Maximum-likelihood estimation of photon-number distribution from homodyne statistics, Phys. Rev. A \textbf{57}, 5013 (1998).

\bibitem{Yuen78}
H. P. Yuen and J. H. Shapiro, Quantum Statistics of Homodyne and Heterodyne Detection, \emph{in Coherence and Quantum Optics IV}, edited by L. Mandel and E. Wolf (Plenum, New York, 1978), p. 719.

\bibitem{Campos89}
R. A. Campos, B. E. A. Saleh, and M. C. Teich, Quantummechanical lossless beam splitter: SU(2) symmetry and photon statistics, Phys. Rev. A \textbf{40}, 1371 (1989).

\bibitem{Leonhardt93}
U. Leonhardt and H. Paul, Realistic optical homodyne measurements and quasiprobability distributions, Phys. Rev. A \textbf{48}, 4598 (1993).

\end{thebibliography}
\end{document}